\documentclass[12pt,a4paper]{iopart}
\usepackage{iopams}
\usepackage{graphicx}
\usepackage[usenames]{color}
\usepackage[normalem]{ulem}

\newcommand{\cpc}{\emph{Comp. Phys. Commun.\ }}

\begin{document} 

\title{Vortex dynamics of rotating dipolar Bose-Einstein condensates}

\author{R. Kishor Kumar and P. Muruganandam}
\address{School of Physics, Bharathidasan University, Palkalaiperur Campus, Tiruchirappalli 620024, Tamilnadu, India}

\begin{abstract}

We study the influence of dipole-dipole interaction  on the formation of vortices in a rotating  dipolar Bose-Einstein condensate (BEC) of $^{52}$Cr and $^{164}$Dy atoms in quasi two-dimensional geometry. By numerically solving the corresponding time-dependent mean-field Gross-Pitaevskii equation, we show that the dipolar interaction enhances the number of vortices while a repulsive contact interaction increases the stability of the vortices. Further, an ordered vortex lattice of relatively large number of vortices is found in a strongly dipolar BEC.

\end{abstract}
\pacs{03.75.Lm, 67.85.De}
\maketitle

\section{Introduction}

The most fundamental properties of a superfluid is irrotational ($\nabla \times v_s=0$, $v_s$ the superfluid velocity) and resistance free flow. An important theoretical development came after the remarkable work by Landau that the excited state of a superfluid would represent the creation of vortices~\cite{Landau1941}. Quantized vortices have long been studied in superconductors, liquid Helium, etc., so called superfluids. The successful experimental realization of Bose-Einstein condensates (BECs) of alkali atoms in the past century has provided a new pathway to many developments in the understanding of the quantized vortices~\cite{Matthews1999, Madison2000, Aboshaeer2001, Pethick, Allan, Fetter2009,Adhikari2002,Bao2006}. In BEC experiments vortices have been nucleated with help of either by laser stirring or by rotating magnetic traps, and they are observed above a certain critical rotation frequency. The observation of quantized vortices could be considered as unquestionable evidence for the existence of superfluidity in BEC.  Quantized vortices in BEC are having obvious analogy with that in liquid Helium and in type-II super conductors~\cite{Donnelly}.

Early experiments and theoretical studies on vortices in BEC have been mostly focused on the conventional Bose gas with local and isotropic interaction. However, many bosonic atoms and molecules have large dipole moments. For example, $^{52}$Cr, $^{164}$Dy,   and $^{168}$Er  BECs, which have a larger anisotropic long-range dipolar interaction superposed on the isotropic short-range atomic interaction, have recently been realized~\cite{Griesmaier2005, Griesmaier2006, Lahaye2007, Koch2008, Lahaye2009, Mingwu2011, Aikawa2012}. The study of dipolar BECs has revealed various interesting properties due to the peculiar competition between an isotropic, short-range contact interaction and an anisotropic, long-range dipolar interaction. These include, the dependence of stability on the trap geometry~\cite{Koch2008,Lahaye2009}, new dispersion relations of elementary excitations~\cite{Ryan2010, Santos2003}, unusual equilibrium shapes, roton-maxon character of the excitation spectrum~\cite{Santos2000, Santos2003, Yi2003, Parker2008} and novel quantum phases including supersolid and checkerboard phases~\cite{Tieleman2011, Zhou2010, Baranov2008}.

Recently, there have been studies on vortices in dipolar BECs using mean-field models~\cite{Baranov2008}. For instance, the structure and stability of vortices in dipolar BECs are found to be strongly affected by the anisotropic character of dipole-dipole interaction (DDI) and relative strengths of the dipolar and contact interactions~\cite{Yi2006,Wilson2009,Abad2009}. Further, static hydrodynamic solutions, and dynamical stability and instability have been explored in a rotating dipolar BEC within Thomas-Fermi limit~\cite{Van2007}. A second-order like phase transition of straight and helical vortex lines influenced by dipolar orientation has also been studied~\cite{Klawunn2008}. More recently the rotational properties of a dipolar BEC in a quasi two-dimensional (2D) geometry for an arbitrary orientation of the dipoles with respect to their plane of motion has been studied~\cite{Malet2011}. 

However, most of the features of vortices in dipolar BECs have not been explored so far. Also there has been no studies on vortices in a strongly dipolar BEC. In the present paper, we analyze the formation of vortices in a dipolar BEC and the influence of strong dipolar interaction on the stability as well as on the number of vortices.  By numerically solving the two-dimensional Gross-Pitaevskii (GP) equation, for a rotating trapped dipolar BEC of $^{52}$Cr and $^{164}$Dy atoms in a quasi-2D geometry, we study formation and dynamics of vortices both in the presence and absence of s-wave (contact) interaction. The number of vortices in a pure dipolar BEC is found to increase when compared to that of a conventional BEC. We compare the vortex number estimated theoretically with that computed from numerical simulations. Further we notice that the inclusion of a repulsive contact interaction in a dipolar BEC enhances the formation of vortex lattices in the dipolar BEC. We also find that the critical rotation frequency decreases with the increase of the strength of dipolar interaction. 

The present paper is organized as follows. In Sec.~\ref{sec:frame} we provide a brief overview on the mean field Gross-Pitaevskii equation describing the properties of a rotating dipolar BEC confined in an axially symmetric harmonic trap potential. A two-dimensional reduction of the GP equation for a dipolar BEC under strong axial confinement is also discussed in Sec.~\ref{sec:frame}. Then, in Sec.~\ref{sec:dyn}, we present the numerical studies on the formation  and dynamics of vortices in dipolar BECs of $^{52}$Cr and $^{164}$Dy atoms. Here we analyze the formation of vortices in a pure dipolar BEC. We also explore the role of contact interaction on the stability, number of vortices and critical rotation frequency. Finally, in Sec.~\ref{sec:conclusion} we provide a summary and conclusion.

\section{Theoretical framework}
\label{sec:frame}

Many phenomenological properties of quantized vortices in BECs can be studied using mean field Gross-Pitaevskii (GP) equation~\cite{Fetter2009, Adhikari2002, Bao2006}. A dipolar BEC with $N$ atoms, each of mass $m$ at absolute zero temperature in a rotating frame can be described by the Gross-Pitaevskii equation as~\cite{Yi2006}
\begin{eqnarray}  
i \frac{\partial \phi({\bf r},t)}{\partial t}  & = & \biggr[ -\frac{1}{2}\nabla^2 +V({\bf r}) 
+ 4\pi a N\vert \phi({\bf r},t)\vert^2 - \Omega L_z \nonumber \\
& & +  N \int U_{dd}({\bf r -r'})\vert\phi({\bf r'},t)\vert^2d^3{ r'}
\biggr] \phi({\bf r},t),  \label{eqn:dgpe} 
\end{eqnarray} 
where $V({\bf r})$ is the confining axially symmetric harmonic potential, $\phi({\bf r},t)$  the wave function at time $t$ with normalization $\int \vert\phi({\bf r},t)\vert^2 d {\bf r}=1$, $a$ the atomic scattering length, which can be tuned to a large extent via Feshbach resonance. The experimental value of the s-wave scattering length of $^{52}\mbox{Cr}$ atom is $a  = (103 \pm 4) a_0$ \cite{Pasquiou2010}.  For the $^{164}\mbox{Dy}$ atoms the s-wave scattering length, yet to be measured, is assumed to be equal to that of $^{52}\mbox{Cr}$~\cite{Mingwu2011}. The axial and radial trap frequencies  are $\Omega_z\omega$ and $\Omega_\rho\omega$, respectively. In equation (\ref{eqn:dgpe}) length is measured in units of characteristic harmonic oscillator length $l \equiv \sqrt{\hbar/m\omega}$, frequency in units of $\omega$, time $t$ in units of $\omega^{-1}$. $L_z= - i (x\partial_y-y \partial_x)$ corresponds to the $z$-component of the angular momentum due to the rotation of the dipolar BEC about $z$ axis with frequency $\Omega$. Here the  $\Omega$ is expressed in units of the radial trap frequency $\Omega_\rho\omega$. The integral term in equation~(\ref{eqn:dgpe}) accounts for the dipole-dipole interaction with
\begin{eqnarray}
U_{dd}({\mathbf  x})=a_{dd}\frac{1-3\cos^2 \theta}{\vert {\mathbf x}\vert ^3},
\end{eqnarray}
where ${\mathbf x}= {\mathbf r} -{\mathbf r'}$ determines the relative position of dipoles and $\theta$ is the angle between ${\mathbf x}$ and the direction of polarization, $z$. 
The constant $a_{dd}=\mu_0\bar{\mu}^2 m/(12\pi\hbar^2)$  is a length characterizing the strength of dipolar interaction and, its experimental value for $^{52}$Cr and $^{164}$Dy are $15a_0$ and $130 a_0$, respectively \cite{Youn2010}, with $a_0$ being the Bohr radius. $\bar{\mu}$ corresponds to the magnetic dipole moment of a single atom and $\mu_0$ the permeability of free space.  

The dimensionless three-dimensional (3D) harmonic trap is given by
\begin{eqnarray}
 V({\bf r}) =  \frac{1}{2}\Omega_\rho^2 \rho^2 + \frac{1}{2} \Omega_z^2 z^2,
\end{eqnarray} 
where ${\bf r}\equiv (\vec\rho,z)$, with $\vec\rho$ the radial coordinate and $z$ the axial coordinate.
In a disk-shape, with strong axial trap frequency ($\Omega_z>\Omega_\rho$), the dipolar BEC is assumed to be in the ground state, 
\begin{eqnarray}
\phi_{1D}(z) = \frac{1}{(\pi d_z^2)^{1/4}}\exp\left(-\frac{z^2}{2d_z^2}\right), \label{eqn:zfun}
\end{eqnarray}
of the axial trap so that the wave function $\phi(\mathbf r)$ can be written as, 
\begin{eqnarray}
\phi(\mathbf r) = \phi_{1D}(z)  \phi_{2D}(x,y),\label{eqn:ans}
\end{eqnarray}
where $\phi_{2D}(x,y)$ is the 2D wave function, $d_{\rho}=l$ and $\lambda d_z^2=1$ with $\lambda = \Omega_z/\Omega_\rho$ the trap aspect ratio. Using ansatz~(\ref{eqn:ans}) in equation~(\ref{eqn:dgpe}), the $z$ dependence can be integrated out to obtain the following effective 2D equation~\cite{Pedri2005,Muruganandam2012,Fisch2006},
\begin{eqnarray}
i\frac{\partial \phi_{2D}(\vec \rho,t)}{\partial t} & = & \biggr[-\frac{\nabla_\rho^2}{2}
+ V_{2D} -\Omega L_z +\frac{4\pi aN}{\sqrt{2\pi}d_z}\vert \phi_{2D}(\vec \rho,t)\vert ^2 \nonumber \\
& & + \frac{4\pi a_{dd}N}{\sqrt{2\pi}d_z} \int \frac{d^2k_{\rho}}{(2\pi)^2} \mbox{e}^{i\bf {k_\rho} \cdot \vec \rho} \, 
{\tilde n}({\bf k_\rho})h_{2D}\left(\frac{k_\rho d_z}{\sqrt{2}} \right) \biggr]  
\phi_{2D}\left(\vec \rho,t \right). \label{red2d}
\end{eqnarray}
where $V_{2D} = [(1+\epsilon)x^2+(1-\epsilon)y^2]/2$ is the trap potential, $\epsilon$ is a parameter used to introduce a small anisotropy for the nucleation of the vortices~\cite{Tsubota2002,Kasamatsu2005}. In equation~(\ref{red2d}),
\begin{eqnarray}
\tilde n({\bf k_\rho}) = \int \exp(i{\bf k_\rho. \vec\rho)} \vert\phi_{2D}(\vec\rho)\vert^2 d\vec\rho,
\end{eqnarray}
$k_\rho\equiv (k_x,k_y)$, $h_{2D}(\xi)=2-3\sqrt{\pi}\xi \exp (\xi^2)\, \mbox{erfc}(\xi)$, and the dipolar term is written in Fourier space. 

In the following, we study the vortices of a disk-shaped dipolar BEC by solving the above equation~(\ref{red2d}) with the aid of a combined split step Crank-Nicolson and Fast Fourier Transform (FFT) based numerical scheme~\cite{Muruganandam2012,Muruganandam2009}. The numerical simulations are carried out with $dx=dy=0.2$ (space step) and $dt=0.005$ (time step).

\section{Formation and dynamics of vortices in dipolar BEC}
\label{sec:dyn}

Before studying the reduced GP equation (\ref{red2d}), it is worth to comment on the validity of 2D approximation, at least, for the trap aspect ratios $\lambda = 30$ and $100$, which we are using throughout this paper. 
\begin{figure}[!ht]
\begin{center}
\includegraphics[width=0.9\columnwidth,clip]{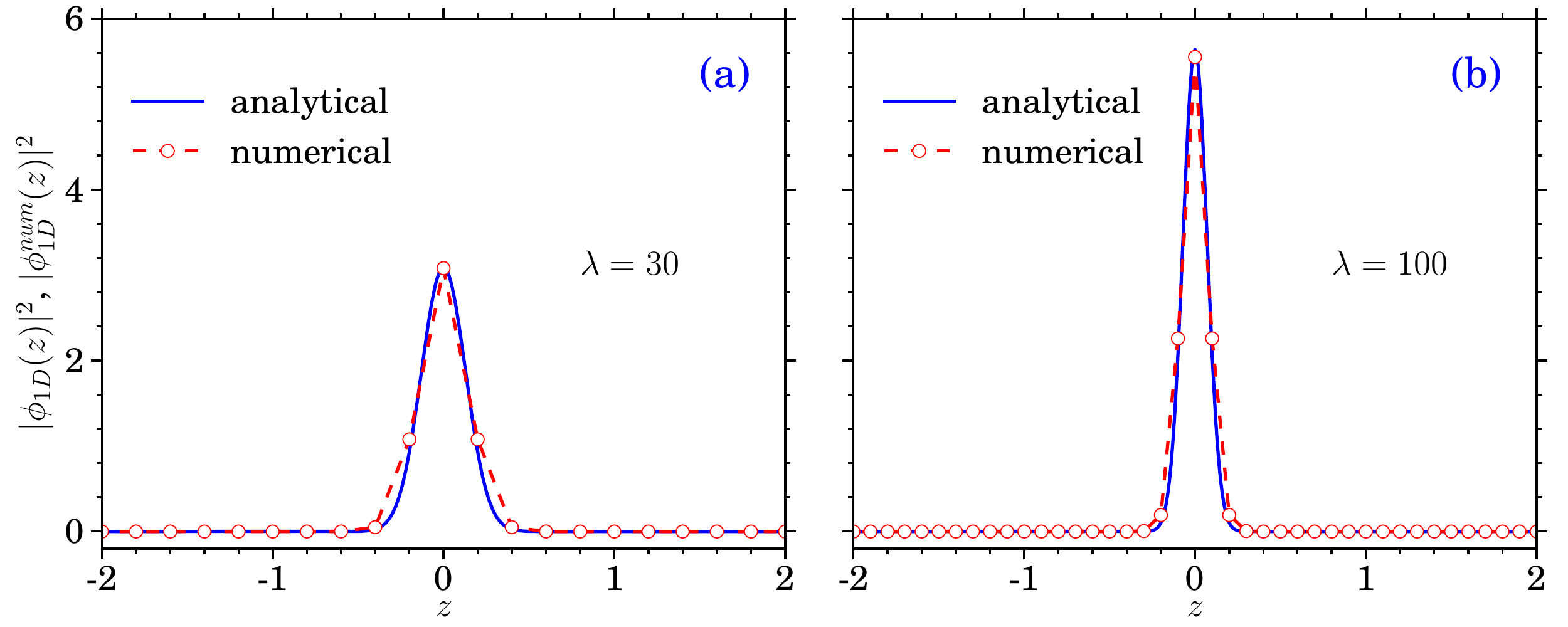}
\end{center}
\caption{(Color online) One-dimensional density profile of a disk-shaped $^{52}$Cr dipolar BEC, of $10\,000$ atoms from 1D wave function (\ref{eqn:zfun}), $\vert \phi_{1D}(z) \vert^2$ and numerical solution of the 3D GP equation (\ref{eqn:dgpe}) with $\Omega = 0$, $\vert \phi_{1D}^{num}(z) \vert^2 $, with trap aspect (a) $\lambda = 30$ and (b)  $\lambda = 100$. The numerical solution of (\ref{eqn:dgpe}) is obtained using imaginary time propagation with (a) $dx=dy=dz=0.2$, $dt = 0.005$ and (b) $dx = dy = dz = 0.1$, $dt = 0.0025$.}
\label{fig:density}
\end{figure}
For this purpose, we compare the one-dimensional density profile of (\ref{eqn:zfun}) and that from the numerical solution of full 3D equation (\ref{eqn:dgpe}) when $\Omega = 0$. The numerical density profile is obtained by integrating out the $x$ and $y$ dependence of the full 3D density,
\begin{eqnarray}
\vert \phi_{1D}^{num}(z) \vert^2 = \int_{-\infty}^{\infty}  \int_{-\infty}^{\infty} \vert \phi(x,y,z)  \vert^2 dx\, dy.
\end{eqnarray}
In Figure \ref{fig:density} we plot the analytical and numerical density profiles  of a disk-shaped $^{52}$Cr dipolar BEC, of $10\,000$ atoms for $\lambda = 30$ and $100$. It is easy to see that the 1D density profiles match very well.

We prepare the ground state wave function by solving equation~(\ref{red2d}) numerically using imaginary time propagation in the absence of angular momentum ($\Omega=0$) and trap anisotropy ($\epsilon=0$). The vortices are then created by evolving the ground state with the inclusion of angular momentum ($\Omega \ne 0$) in real time propagation up to $t \sim 2 \times 10^{4}\,\omega^{-1}$. A small anisotropy, $\epsilon = 0.06$ and a phenomenological dissipation ($\sim 10^{-5}$) are introduced to facilitate the smooth vortex formation~\cite{Tsubota2002,Kasamatsu2005,Garcia-Ripoll2001}. The presence of dissipative mechanism is evident in BEC experiments exhibiting collective damped oscillations~\cite{Tsubota2002}.  Theoretical predictions based on mean-field GP equation with dissipative term found to give correct description of the damping. In a rotating BEC, vortex lattices will never be formed without dissipation even if the trapping potential is rotated fast enough. This phenomenological dissipation is effected by replacing the term `$i\partial/\partial t$' with `$(i-\gamma)\partial/\partial t$' in the time dependent equation (\ref{red2d}) and $\gamma$ accounts for the dissipation. Vortices are formed for rotation frequencies above a critical value, $\Omega > \Omega_c$. For smaller trap aspect ratios, a BEC with dominant dipolar interaction with the dipoles aligned along the trap symmetry axis becomes unstable~\cite{Koch2008,Ryan2009}. Here, we consider a dipolar BEC in disk-shape geometry, $\lambda \gg 1$, as it is more stable. To explore the effects of dipolar interaction on the vortex formation we consider a disk-shape geometry with the number of atoms fixed to $N = 10\, 000$. 

\subsection{Vortices in a pure dipolar BEC}

First, we consider the case of a pure dipolar BEC with $a_{dd} = 15\, a_0$ and $a=0$ with the trap aspect ratio, $\lambda=30$. The formation of vortex is analyzed by varying the rotation frequency $\Omega$ in the range $0.1$ to $0.9$. When the condensate begins to rotate it elongates and the boundary surface becomes unstable.
\begin{figure}[!ht]
\begin{center}
\includegraphics[width=0.8\columnwidth,clip]{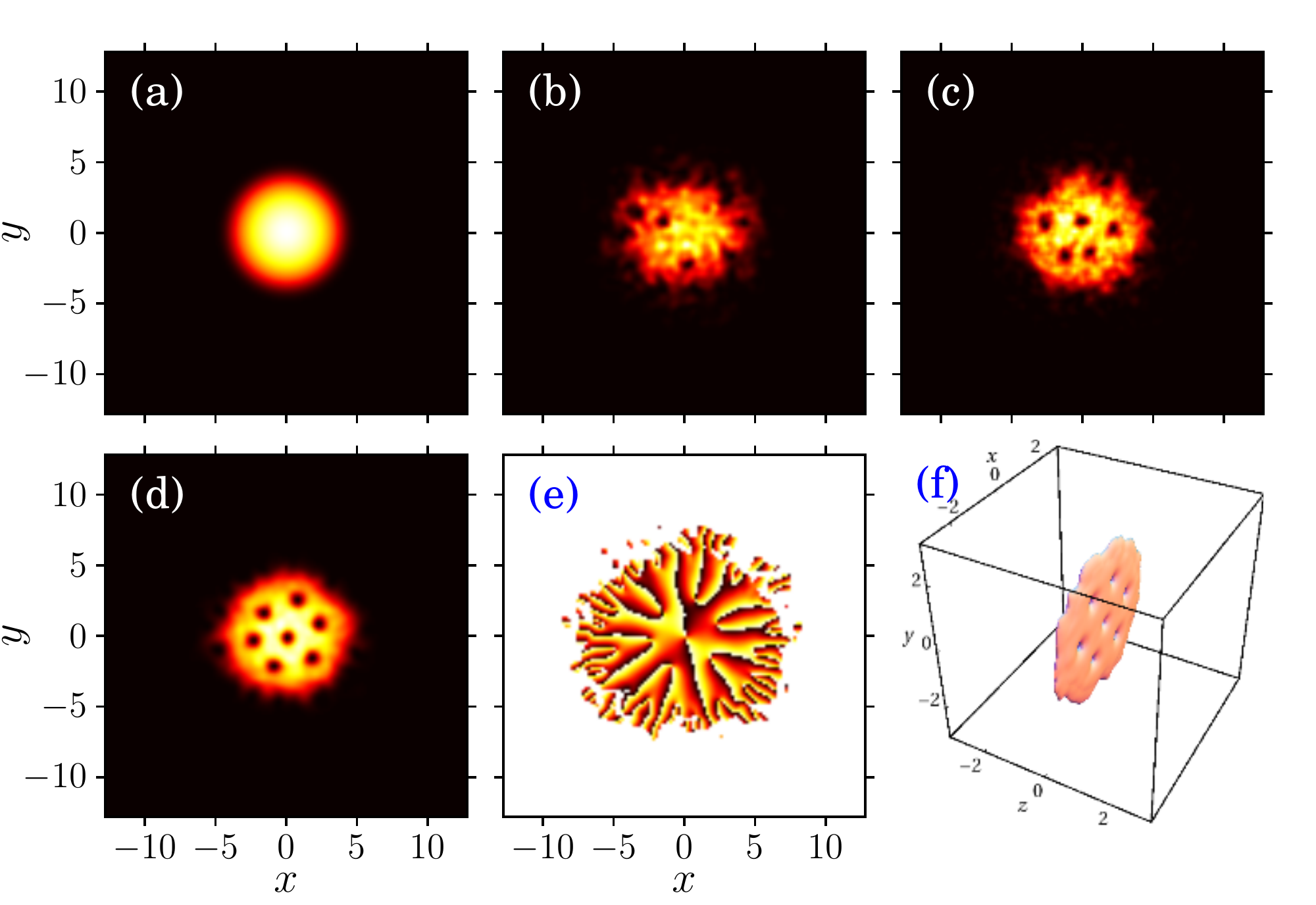}
\end{center}
\caption{(Color online) Contour plots of the density $\vert \phi_{2D}(\rho, t) \vert^2$ showing the development of vortices in a dipolar BEC with $a = 0$, $a_{dd} = 15\, a_0$, $N = 10\,000$, $\lambda = 30$ and $\Omega = 0.6$ after the trapping potential begins to rotate: (a) $t=0$ (b) $t=1000\, \omega^{-1}$,  (c) $t=2000\, \omega^{-1}$, (d) $t = 20\, 000\, \omega^{-1}$, (e) Phase distribution of condensate wave function $\phi_{2D}(\rho, t)$  of fully developed vortices (steady vortex state) shown in (d), and (f) Three dimensional contour plot showing the vortices obtained from the numerical simulation of the full 3D equation (\ref{eqn:dgpe}) with the above parameters.}
\label{fig1}
\end{figure}
Then ripples are developed on the surface and as a consequence vortices enter into the condensate above a critical rotation frequency, $\Omega_c \sim 0.38 $. As time progress these vortices approach to a stable configuration. The density plots of $\vert \phi_{2D}(\rho, t) \vert^2$ of the time development of vortices for $\Omega=0.6$ are shown in Figures~\ref{fig1}(a)-(d). About 7 vortices form a stable pattern as shown in Figure~\ref{fig1}(d). We have calculated the phase from the final wave function $\phi(\vec\rho, t)$, which varies continuously from $0$ (dark) to $2\pi$ (bright) and shown in Figure~\ref{fig1}(e). At the place where a vortex is located, a bifurcation (branching) in the phase is visible.

We have also performed the numerical simulation using the full 3D equation (\ref{eqn:dgpe}) for the above set of parameters with the same rotation frequency, which again confirms the number of vortices and is shown in Figure~\ref{fig1}(f).  
\begin{figure}[!ht]
\begin{center}
\includegraphics[width=0.6\columnwidth,clip]{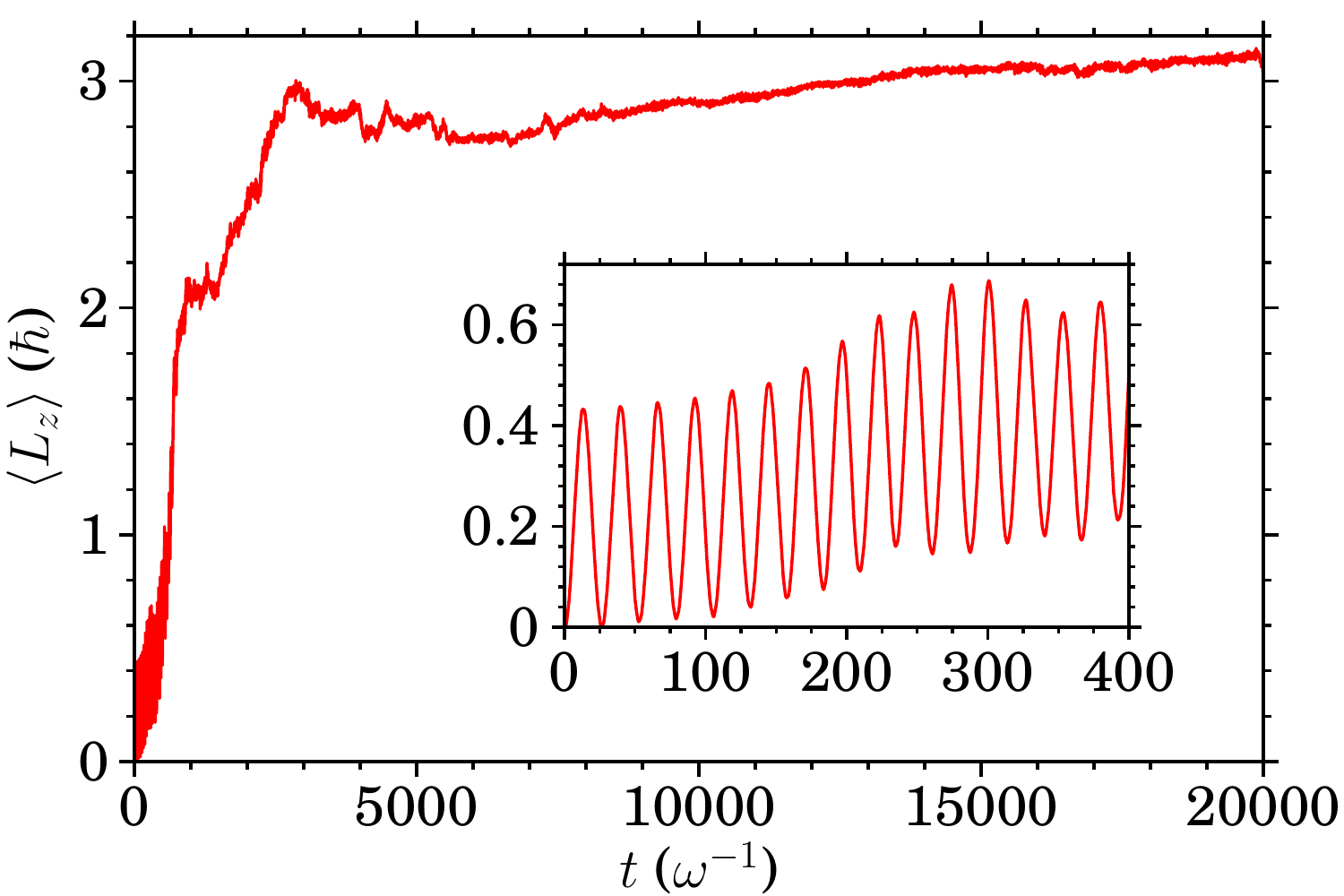}
\end{center}
\caption{(Color online) Time evolution of the expectation value of angular momentum $\langle L_z \rangle$ during the development of vortices shown in Figure~\ref{fig1}}
\label{fig-ang}
\end{figure} 
Further, the expectation value of angular momentum, defined as
\begin{eqnarray}
\langle L_z \rangle = i \int \phi^{\star}(\vec \rho, t) (y\partial_x-x\partial_y) \phi(\vec \rho, t)\, d\vec\rho,
\end{eqnarray}
is also calculated as a function of time and is shown in Figure~\ref{fig-ang}. The expectation value of angular momentum $\langle L_z \rangle$ gradually increases with periodic oscillation (see inset in Figure~\ref{fig-ang}) and then settles to almost steady value confirming the stable vortex pattern. The oscillation is due to the shape (quadrupole) deformation of the condensate at vortex free state and it is associated with creation of vortices. 
\begin{figure}[!ht]
\begin{center}
\includegraphics[width=0.8\columnwidth,clip]{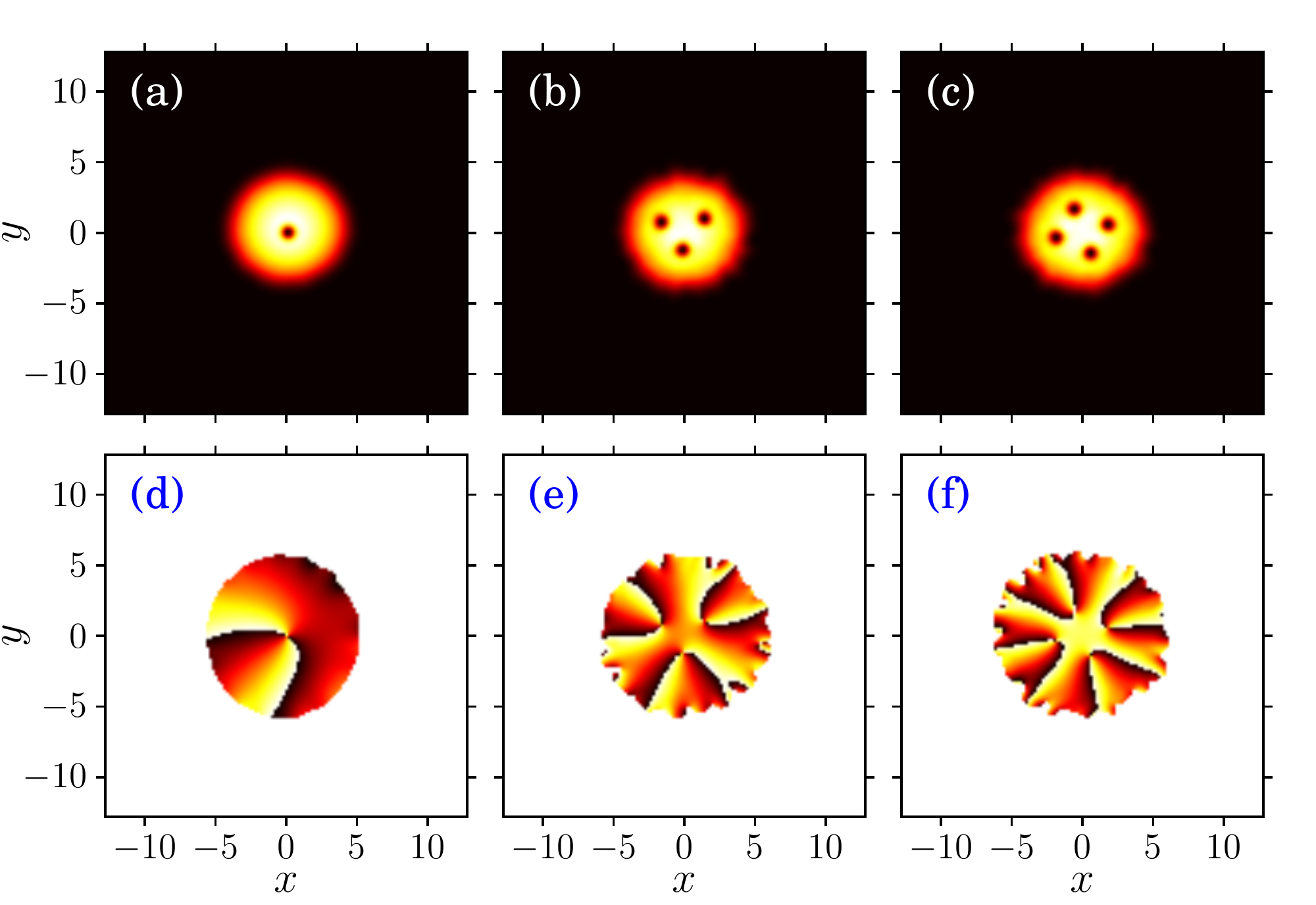}
\end{center}
\caption{(Color online) Contour plots of the density distributions $\vert \phi_{2D} \vert^2$ showing steady vortex state in a rotating dipolar BEC with $a_{dd}=15 \,a_0$, $a=0$, $N = 10\,000$, $\lambda = 30$: (a) $\Omega \equiv \Omega_c= 0.38$, (b)  $\Omega = 0.44$, and (c) $\Omega = 0.47$. (d) - (f) The corresponding phase distributions of the condensate wave function $\phi_{2D}$ of (a) - (c).}
\label{fig2}
\end{figure}
In Figures~\ref{fig2}(a)-(c) and \ref{fig2}(d)-(f) we depict the contour plots of the density $\vert \phi_{2D}\vert^2$ and the corresponding phase distribution of condensate wave function $\phi_{2D}$ for different rotation frequencies, $\Omega = 0.38$, $0.44$, and $0.47$, respectively.

Next we consider the development of vortices in a dipolar BEC with larger dipolar strength, for example, $a_{dd} = 130\, a_0$. In this case, the critical rotation frequency is found to be $\Omega_c=0.271$, which is small when compared to that observed with $a_{dd} = 15\, a_0$ above.  Thus, a BEC with larger dipolar interaction nucleates vortices with smaller rotation frequency~\cite{Abad2009}.
\begin{figure}[!ht]
\begin{center}
\includegraphics[width=0.8\columnwidth,clip]{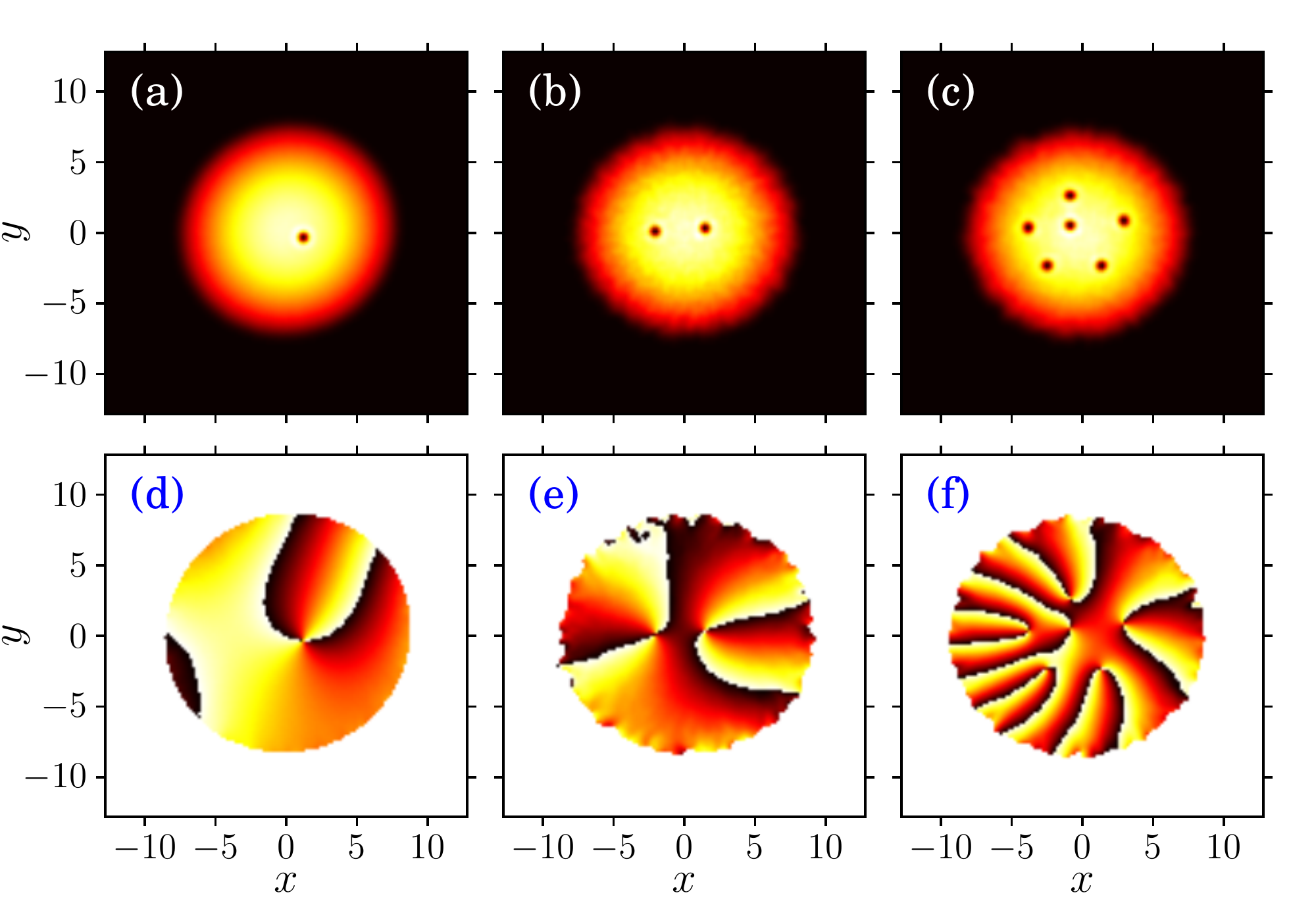}
\end{center}
\caption{(Color online) Contour plots of the  density distributions $\vert \phi_{2D} \vert^2$ showing steady vortices state in a rotating dipolar BEC with $a_{dd}=130 \,a_0$, $a=0$, $N = 10\,000$, $\lambda = 30$: (a) $\Omega \equiv \Omega_c= 0.271$, (b)  $\Omega = 0.29$, and (c) $\Omega = 0.33$. (d) - (f) The corresponding phase distributions of the condensate wave function $\phi_{2D}$ of (a) - (c).}
\label{fig3}
\end{figure}
At critical rotation frequency, $\Omega_c$ a single vortex is formed as shown Figure~\ref{fig3}(a), while the corresponding phase is depicted Figure~\ref{fig3}(d). 
On increasing $\Omega$ above $\Omega_c$ more number of vortices are created as illustrated in Figures~\ref{fig3}(b)-(c) for $\Omega = 0.29$ and $\Omega = 0.33$, respectively, and the corresponding phases are shown in Figures~\ref{fig3}(d)-(f).

Further, it is interesting to note that in Figure~\ref{fig3}, vortex arrays as well as a single vortex, are not centered with respect to the trap symmetry axis, in contrast to the centered vortex arrays that are shown in Figure~\ref{fig2} for a smaller dipolar interaction strength as it happens also in conventional BECs. In this case, the quadruple oscillation persists for longer times. This is essentially due to the strong dipolar strength, $a_{dd} = 130\, a_0$, of $^{164}$Dy atoms. 

On increasing $\Omega$ further, the system with $a_{dd} = 15\, a_0$ destabilizes as $\Omega$ approaches unity. Moreover, a pure dipolar BEC with $a_{dd} = 130\, a_0$ becomes unstable for $\Omega > 0.62$. Rotation frequencies larger than $0.62$ creates a large number of vortices and makes the boundary surfaces more unstable. Certain peculiar dynamics is observed during the formation of vortices when the rotation frequency $\Omega \lesssim 0.5$ for larger dipolar strength ($a_{dd} = 130\, a_0$). The dipolar BEC deforms as it rotates and the boundary surfaces become unstable, as a result more vortices enter and remains stable for a while. Then,
\begin{figure}[!ht]
\begin{center}
\includegraphics[width=0.8\columnwidth,clip]{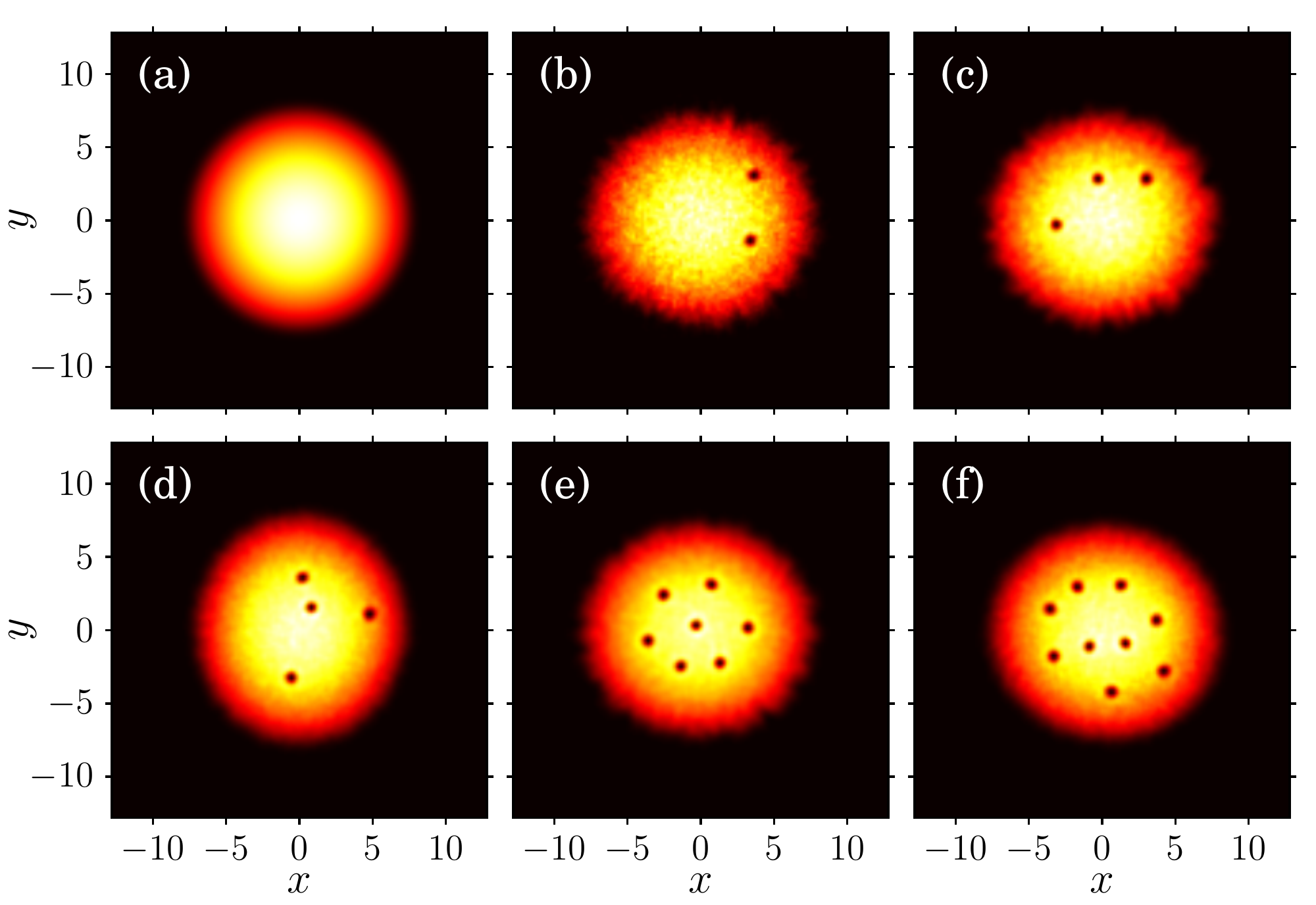}
\end{center}
\caption{(Color online) Contour plots of the density $\vert \phi_{2D}(\vec \rho, t)\vert^2$ showing vortices in a rotating dipolar BEC with $a \, = \,0$, $a_{dd}=130\,a_0$, $N = 10\,000$, $\lambda = 30$ and $\Omega = 0.35$: (a) $t=0$ (b) $t=30\,000\, \omega^{-1}$,  (c) $t=40\,000\, \omega^{-1}$, (d) $t=50\,000\, \omega^{-1}$, (e) $t = 80\,000\, \omega^{-1}$, and (f) $t = 100\,000\, \omega^{-1}$.}
\label{figdyn}
\end{figure}
as time progress, again the surfaces become unstable and few more vortices enter into the condensate and this scenario repeats for reasonably longer times. In this case $\langle L_z \rangle$ shows periodic oscillations for a very long time. In Figure~\ref{figdyn} we plot the snapshots of density $\vert \phi_{2D}\vert^2$ showing the dynamics. Further, it is evident that the number of vortices increasing rather slowly as time progress. However, the above dynamics is not seen when the dipolar strength is relatively low or zero.

We also examine the speed at which the vortices are formed in a dipolar BEC. In conventional BECs, with repulsive contact interaction, it takes about few hundred milliseconds for the propagation of surface ripples that are created due to trap deformation to the vortex formation~\cite{Tsubota2002,Kasamatsu2005}. While in dipolar BECs, apart from trap deformation, the anisotropy nature of DDI breaks the axial symmetry more easily~\cite{Malet2011} and stimulates rapid vortex formation. Our numerical simulations also confirm this rapid vortex formation as it take about few ten milliseconds for the creation of vortices, which is one order less compared to non-dipolar BECs. We considered a similar set of trap parameters, contact interaction strength and rotation frequency as in Ref.~\cite{Tsubota2002}.

Next, it is of interest to study the dependence of the vortex number ($N_v$) on the rotation frequency $\Omega$ in dipolar BEC. In a rotating superfluid, the number of vortices formed for a given rotation frequency can be predicted using the Feynman's rule~\cite{Feyn1955}, that is, $N_v=({m\Omega}/{\hbar})R_\rho^2(\Omega)$, where $R_\rho(\Omega)$ is the radius of the rotating fluid. In the Thomas-Fermi (TF) limit, the radius of the trapped BEC under rotation is given by \cite{Fetter2009},
\begin{eqnarray}
R_\rho(\Omega) = R_\rho(0) \left(1-\frac{\Omega^2}{\Omega_\rho^2}\right)^{-1/{4}}.\label{feyn1}
\end{eqnarray}
Then the relation connecting vortex number and the rotation frequency can be written as
\begin{eqnarray}
N_v = \Omega \, R_\rho^2(0) \left(1-\frac{\Omega^2}{\Omega_\rho^2}\right)^{-1/2}.\label{feyn2}
\end{eqnarray}
The Thomas-Fermi (3D disk/pancake) radius $R_{TF}(0)$ is given by \cite{Parker2008}
\begin{eqnarray}
R_{TF}(0) = \left[ 15N  \lambda  (a + 2 a_{dd}) \right]^{1/5}.
\end{eqnarray}
Here we use $R_\rho(0) =  R_{TF}(0)/\sqrt{3}$~\cite{Abad2010}.

The above relation (\ref{feyn1}) is used to study the vortex number in conventional BECs in a rotating deep optical lattice~\cite{Kato2011}. Here we use the above equation~(\ref{feyn2}) to estimate the vortex number in a rotating dipolar BEC. This will be more useful to confirm the dominance of dipolar strength over Feynman's rule. We also calculate the vortex number as a function of the rotation frequency $\Omega$ by numerically solving the the GP equation (\ref{red2d}). The numerical vortex numbers are calculated in the equilibrium state by evolving equation~(\ref{red2d}) in real time propagation for sufficiently longer times, typically of the order of $150\,000 \, \omega^{-1}$ to $200\,000 \, \omega^{-1}$. 

\begin{figure}[!ht]
\begin{center}
\includegraphics[width=\columnwidth,clip]{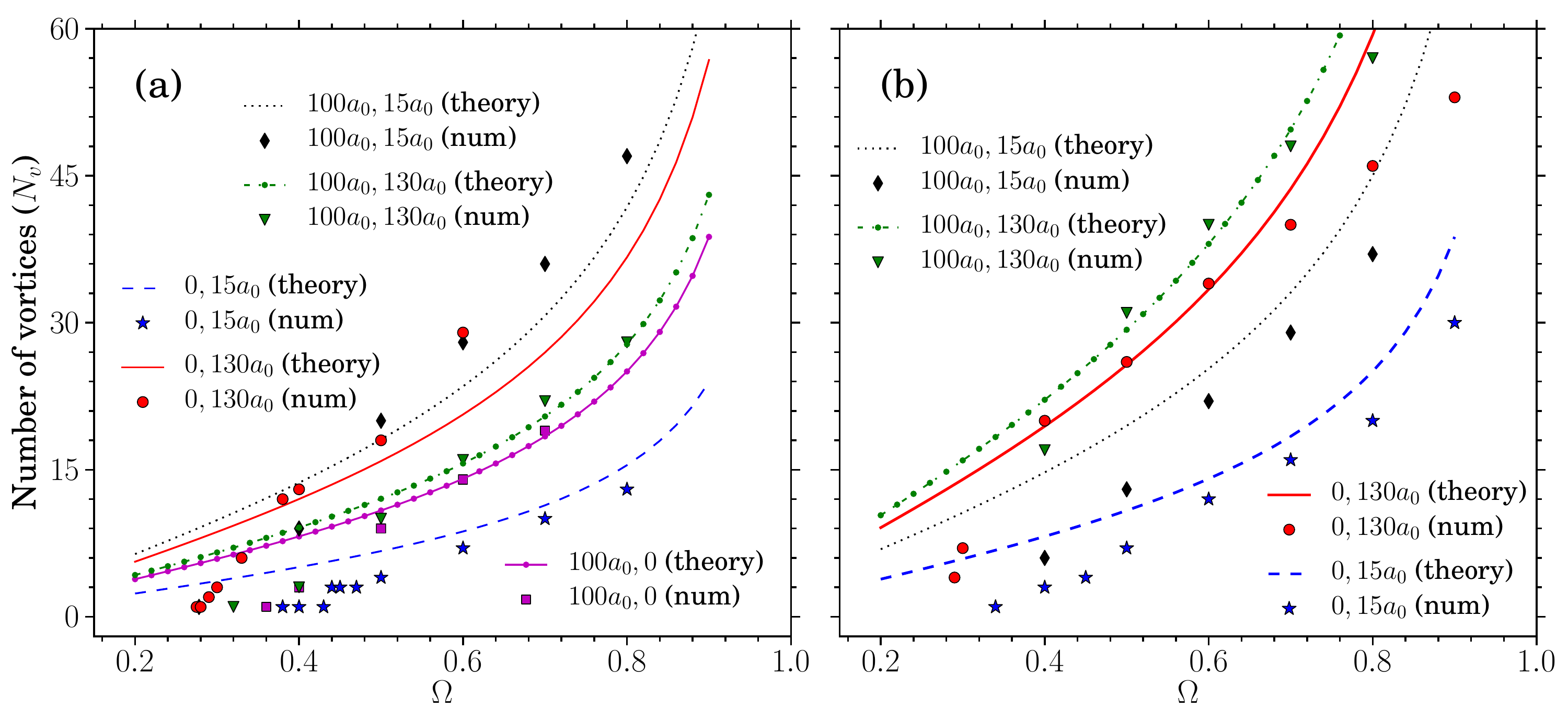}
\end{center}
\caption{(Color online) Plot of the equilibrium vortex number ($N_v$) as a function of the rotation frequency $\Omega$ for conventional BEC with $a = 100 a_0$, dipolar BECs with $a_{dd} = 15\, a_0$ ($^{52}$Cr) and $a_{dd} = 130\, a_0$ ($^{164}$Dy), and  $a = 0$ (pure dipolar) as well as $a = 100 a_0$ for different trap aspect ratios: (a) $\lambda = 30$ and (b) $\lambda = 100$. Here `theory' represents the vortex number estimated using equation~(\ref{feyn2}) and `num' corresponds to numerically computed values.} 
\label{fig-nv}
\end{figure}

In Figure~\ref{fig-nv}(a) we plot the vortex number $N_v$ obtained by using equation~(\ref{feyn2}) (theory) and that calculated numerically (num) for $a=0$, $a_{dd} = 15 a_0$ and $130 a_0$ for $\lambda=30$. We also show the vortex number for the case of conventional BEC with $a \neq 0$ and $a_{dd} = 0$ and dipolar BEC with $a=100 a_0$.
For $a_{dd} = 15 a_0$ the numerical and theoretical vortex numbers agree reasonably well.However, for larger dipolar strength, $a_{dd} = 130 a_0$, with higher rotation frequencies, $\Omega \geq 0.6$ the numerical results deviates from results of Thomas-Fermi approximation. The reason is due to the fact that for stronger dipolar interaction with larger rotation frequencies, the shape of the density profile may not be correctly described by the parabolic Thomas-Fermi profile.

We also study the vortex number of a dipolar BEC confined in a harmonic trap with large trap aspect ratio, for example, $\lambda = 100$. Figure~\ref{fig-nv}(b) shows the plots of numerical and theoretical vortex number $N_v$ with $\Omega$ for a dipolar BECs with $a=0$, $a_{dd} = 15 a_0$ and $130 a_0$ for $\lambda=100$. We find a similar dependence of the vortex number on the rotation frequency as above. However, the rotating dipolar BEC ($a_{dd} = 130\, a_0$) is stable for larger rotation frequencies as against $\Omega < 0.62$ seen earlier with $\lambda = 30$.

\subsection{Effect of contact interaction on vortices}

In the above, we have mostly analyzed the formation of vortices in a dipolar BEC with contact interaction set to zero. In general, the presence of a repulsive contact interaction enhances the stability of the dipolar BEC~\cite{Koch2008}. It is of interest to study the role of inclusion of contact interaction $a$ on the formation of vortices. 
In order to see the effect of inclusion of contact interaction on the formation and stability of vortices we consider the case of rotating dipolar BEC with suitable repulsive contact interaction.

First, we fix the contact interaction strength, say for example $a = 10\, a_0$, and vary the $\Omega$ to analyze the vortex formation in $^{52}$Cr and $^{164}$Dy condensates.
\begin{figure}[!ht]
\begin{center}
\includegraphics[width=0.8\columnwidth,clip]{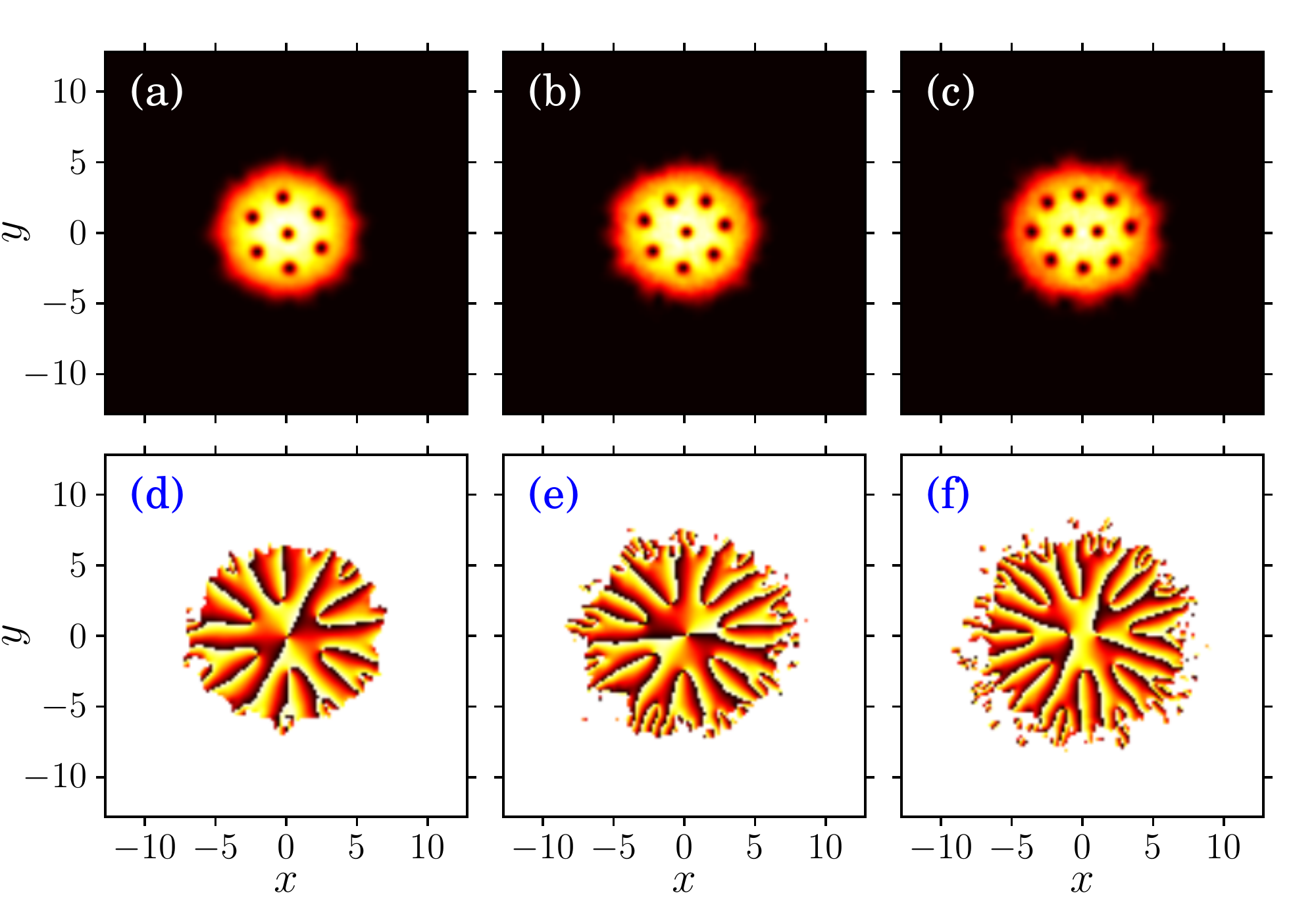}
\end{center}
\caption{(Color online) Contour plots of the density distributions $\vert \phi_{2D}\vert^2$ of a rotating dipolar BEC  showing steady vortex state with $a = \,10 a_0$, $a_{dd}=15 \,a_0$, $N = 10\,000$, $\lambda = 30$: (a) $\Omega = 0.5$ (b) $\Omega = 0.6$ and (c) $\Omega =0.65$.  (d) - (f) The corresponding phase distributions of the condensate wave function $\phi_{2D}$ of (a) - (c).}
\label{fig6}
\end{figure}
In Figures~\ref{fig6}(a)-(c) we plot the density profiles of the vortices with dipolar strengths $a_{dd} = 15\, a_0$ ($^{52}$Cr), for different rotation frequencies $\Omega = 0.5$, $0.6$ and $0.65$, respectively. The corresponding phase distribution of the condensate wave function is plotted in Figures~\ref{fig6}(d)-(f). For instance, a stable vortex lattice with eight vortices as shown in Figure~\ref{fig6}(b) is found for $a_{dd} = 15\, a_0$ and $\Omega = 0.6$ with $a = 10\, a_0$. While in the absence of contact interaction it supports only $7$ vortices as seen in Figure~\ref{fig1}(e). It is evident that the inclusion of repulsive contact interaction in a dipolar BEC increases the number of vortices. 

\begin{figure}[!ht]
\begin{center}
\includegraphics[width=0.8\columnwidth,clip]{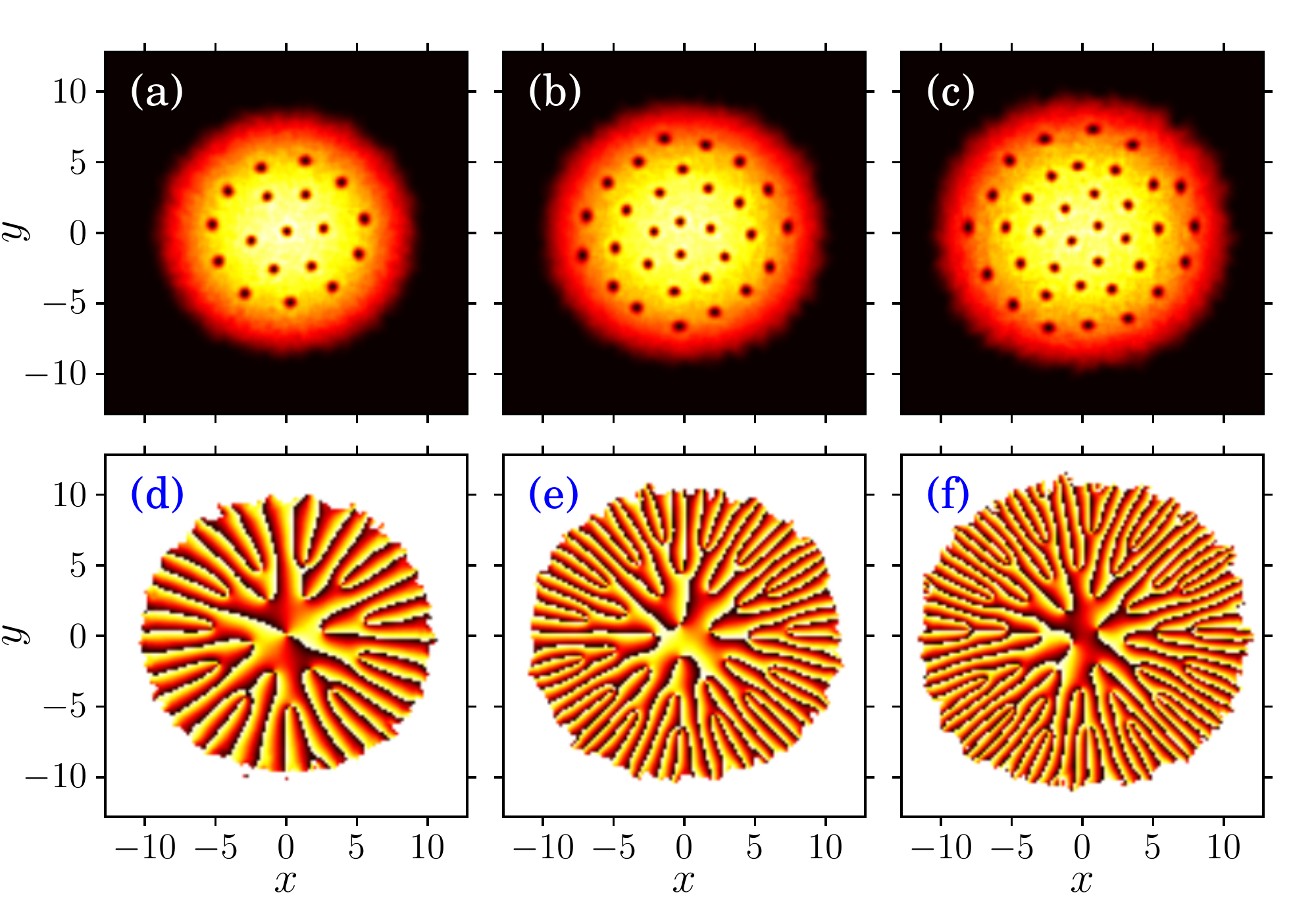}
\end{center}
\caption{(Color online) Contour plots of the density distributions $\vert \phi_{2D}\vert^2$ of a rotating dipolar BEC  showing   steady vortex state with $a = 10\, a_0$, $a_{dd}=130 \,a_0$, $N = 10\,000$, $\lambda = 30$: (a) $\Omega  = 0.5$ (b) $\Omega  = 0.6$ and (c) $\Omega =0.65$.  (d) - (f) The corresponding phase distributions of the condensate wave function $\phi_{2D}$ of (a) - (c).}
\label{fig7}
\end{figure}

In Figures~\ref{fig7}(a) - (c) we show the contour plots of $\vert \phi_{2D}\vert^2$ of a stronger dipolar BEC with and $a = 10 a_0$  and $a_{dd} = 130 a_0$  ($^{164}$Dy) for rotation frequencies $\Omega = 0.5$, $0.6$ and $0.65$. Figures~\ref{fig7}(d) - (f) depict the corresponding phase distributions of $\phi_{2D}$. 
As many as $18$, $30$ and $35$, vortices for $\Omega = 0.5$, $0.6$ and $0.65$, respectively [Figures~\ref{fig7}(a) - (c)] has been observed.  We saw earlier that a dipolar BEC with $a=0$ and $a_{dd}=130 a_0$ becomes unstable for rotation frequencies, $\Omega > 0.62$. Thus, as expected \cite{Koch2008}, the inclusion of a small repulsive contact interaction ($a = 10 a_0$) enhances the stability of the rotating dipolar BEC. 

\begin{figure}[!ht]
\begin{center}
\includegraphics[width=0.8\columnwidth,clip]{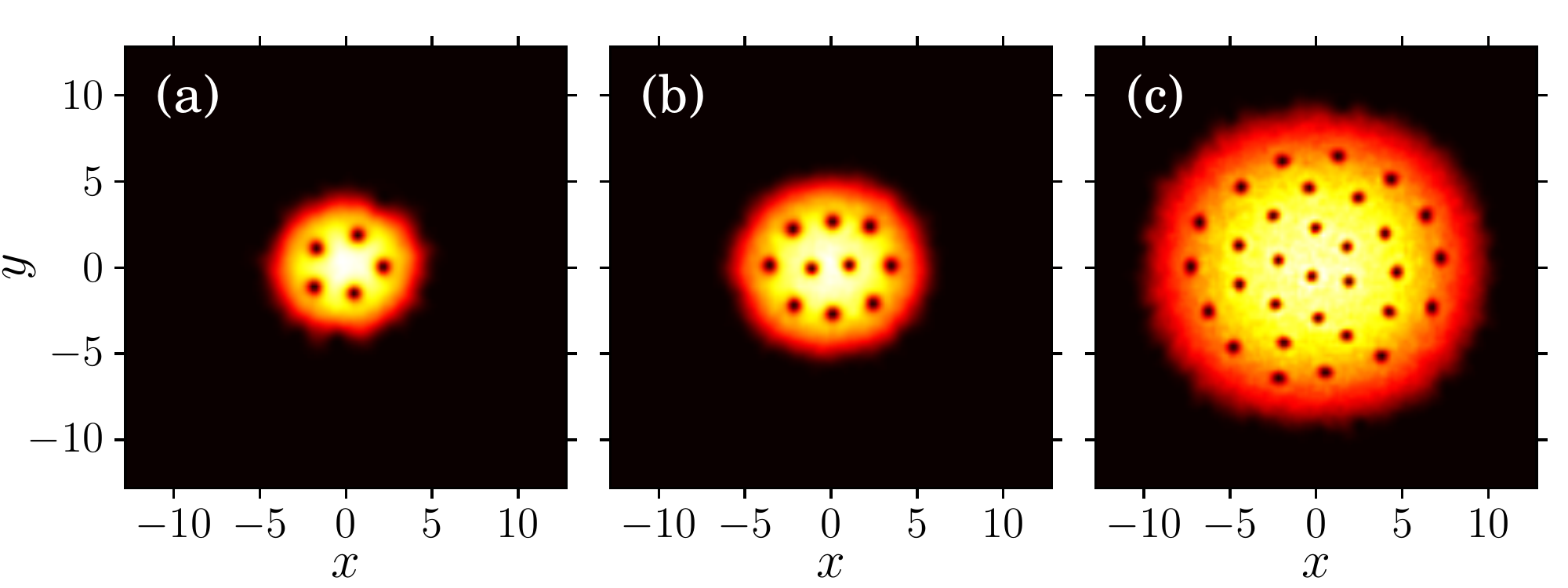}
\end{center}
\caption{(Color online) Contour plots of the density distributions $\vert \phi_{2D}\vert^2$ of a rotating dipolar BEC showing  steady vortex state with  $a = \,20 a_0$, $\Omega=0.6$, $N=10000$, $\lambda=30$: (a) $a_{dd}=0$ (b) $a_{dd}= 15\, a_0$ and (c) $a_{dd}=130\,a_0$.}
\label{fig8}
\end{figure}

Next we  study the effect of dipolar strength in the formation of vortices for a fixed contact interaction strength and rotation frequency. By choosing the condensate parameters as $a=20 \,a_0$, $\Omega=0.6$, and $\lambda=30$ we analyze the pattern of vortices for different dipolar strengths. About $5$ vortices were observed in the absence of dipolar interaction, that is, $a_{dd}=0$, as shown in Figure~\ref{fig8}(a). 
This number increases to $10$  when $a_{dd} = 15 a_0$ [Figure~\ref{fig8}(b)] and further to $31$  for $a_{dd}=130 a_0$ [Figure~\ref{fig8}(c)]. This indicates that the dipolar interaction favors more number of vortices in a rotating BEC.
\begin{table}[!ht]
\caption{Critical rotation frequency $\Omega_c$ for vortices in a dipolar BEC}
\label{table1}
\begin{center}
     \begin{tabular*}{0.5\columnwidth}{@{\extracolsep{\fill}}r r c}
        \hline \hline
        \multicolumn{1}{c}{$a/a_0$} 
	  & \multicolumn{1}{c}{$a_{dd}/a_0$} 
	  & \multicolumn{1}{c}{$\Omega_c$}   \\[1mm] \hline
         $0$   &     $15$      &     $0.380$ \\
         $0$   &    $130$      &     $0.271$ \\
        $20$   &      $0$      &     $0.500$ \\ 
        $20$   &     $15$      &     $0.390$ \\
        $20$   &    $130$      &     $0.275$ \\
       $100$   &      $0$      &     $0.360$ \\ 
       $100$   &     $15$      &     $0.320$ \\
        \hline  \hline
    \end{tabular*}
\end{center}
\end{table}

We also calculate the critical rotation frequencies $\Omega_c$ of a dipolar BEC with different contact and dipolar interaction strengths. The results are given in Table~\ref{table1}. The critical rotation frequency, $\Omega_c$, for a rotating dipolar BEC decreases with the increase (or addition) of dipolar interaction or contact interaction. Similar conclusions have been drawn in earlier studies, which suggest that the critical rotation frequency depends on both the strengths of contact as well as dipole-dipole interactions~\cite{Abad2009}. However, the dipolar interaction play more dominant role in decreasing $\Omega_c$ than contact interaction. 

\section{Summary and Conclusion}
\label{sec:conclusion}

We have studied the influence of dipole-dipole interaction on the formation of vortices in a rotating dipolar Bose-Einstein condensate of $^{52}$Cr and $^{164}$Dy atoms. In particular, we have performed numerical simulation of the quasi-2D Gross-Pitaevskii equation and showed that vortices appear more rapidly in a dipolar BEC when compared to non-dipolar BECs. Further, the number of vortices ($N_v$) is found to be significant in a dipolar BEC and increases monotonically with the increase of rotation frequency. We also compared the number of vortices with Feynman rule. Our analysis suggests that for stronger dipole-dipole interaction strengths and higher rotation frequencies $N_v$ falls well outside the curve predicted by Feynman rule. We have calculated the critical rotation frequency ($\Omega_c$) for vortices, which decreases considerably with the increase of dipolar strength.

The competition between the contact and dipolar interaction on the formation of vortices in a dipolar BEC is found to be complementary as the contact interaction enhances the  stability of the vortices whereas the dipolar interaction increases the number of vortices. 

\ack

This work forms a part of Department of Science and Technology (Ref. No. SR/S2/HEP-03/2009) and Council of Scientific and Industrial Research (Ref. No. 03(1186)/10/EMR-II), Government of India funded research projects.

\end{document}